\documentclass[12pt]{iopart}

\usepackage{graphicx}
\usepackage{dcolumn}
\usepackage{bm}
\usepackage{amssymb}
\usepackage{epsf}
\usepackage{color}
\usepackage[utf8]{inputenc}

\begin{document}

\title[....non-equilibrium field cooled state]{Non-equilibrium magnetic response of canonical spin glass and magnetic glass\\}

\author{Sudip Pal, Kranti Kumar, A. Banerjee and S. B. Roy}

\address{%
 UGC DAE Consortium for Scientific Research\\
 Indore-452001, India 
}%
\author{A. K. Nigam}
\address{%
 Tata Institute of Fundamental Research\\
 Mumbai-400005, India 
}%
\date{\today}

\begin{abstract}
Time and history dependent magnetization has been observed in a wide variety of materials, which are collectively termed as the glassy magnetic systems. However, such systems showing similar non-equilibrium magnetic response can be microscopically very different and they can be distinguished by carefully looking into the details of the observed metastable magnetic behavior. Canonical spin glass is the most well studied member of this class and has been extensively investigated both experimentally and theoretically over the last five decades. In canonical spin glasses, the low temperature magnetic state obtained by cooling across the spin glass transition temperature in presence of an applied magnetic field is known as the field cooled (FC) state. This FC state in canonical spin glass is widely believed as an equilibrium state arising out of a thermodynamic second order phase transition. Here, we show that the FC state in canonical spin glass is not really an equilibrium state of the system. We report careful dc magnetization and ac susceptibility  measurements on two canonical spin glass systems, AuMn (1.8\%) and AgMn (1.1\%). The dc magnetization in the FC state shows clear temperature dependence. In addition, the magnetization shows a distinct thermal hysteresis in the temperature regime below the spin glass transition temperature. On the other hand, the temperature dependence of ac susceptibility has clear frequency dispersion below spin glass transition in the FC state prepared by cooling the sample in the presence of a dc-bias field. We further distinguish the metastable response of the FC state of canonical spin glass from the metastable response the FC state in an entirely different class of glassy magnetic system namely magnetic glass, where the non-equilibrium behavior is associated with the kinetic-arrest of a first order magnetic phase transition.  
\end{abstract}
                             

\section{Introduction} 
Over the years, an intriguing class of magnetic systems having competing magnetic interactions and resultant frustration has been a subject of much attention. The dynamical properties of such magnetic systems have considerable similarities with the dynamical properties of structural glasses. The most well studied member of this category of materials is the canonical spin glass (SG). The canonical SG represent the dilute magnetic alloys where minute amount (within few percent) of transition metal (TM) atoms like Mn, Fe etc., are randomly distributed in the matrix of noble metals like Cu, Au, Ag etc., and the localized magnetic moments on TM atoms interact through spatially oscillating Ruderman–Kittel–Kasuya–Yosida (RKKY) interaction. The canonical SGs have been intensely investigated over the last five decades, but a complete understanding of the SG phenomena is still far from being achieved {\color{blue}\cite{RMP1986, Mydosh2015}}. As the concentration of TMs is further increased, it gives rise to more complicated kinds of glass-like magnetic states with short range magnetic order like concentrated SG, cluster glass, mictomagnet etc. Such glass-like magnetic states can also be achieved in variety of other systems with competing magnetic interactions, other than metallic alloys. On the other hand, a distinctly different kind of glassy state is observed when a first order magnetic transition remains incomplete even beyond its supercooling limit in magnetic field (H)- temperature (T) phase space. The materials showing such glass-like magnetic behaviour arising out of the kinetic arrest of a first order phase transition are termed as magnetic glasses {\color{blue}\cite{MKC2003, RS2006, KS2006,AB2006, VKS2007, AB2008,PC2008,AB2009,RS2009,RS2013,EPL2013}}. It is important to note at this point that all such systems namely spin glass, cluster glass and magnetic glass are actually quite distinct in their microscopic ground state, while showing apparently similar non-equilibrium magnetic response. There exist subtle but distinct differences in their dynamical magnetic response too, and this can only be differentiated through a systematic and careful investigation of their metastable magnetic properties {\color{blue}\cite{RS2009, Sudip2019}}. However, such experiments are few and far between, and often observed magnetic properties are rationalized within a generalized mean-field theoretical framework of SG {\color{blue}\cite{Nayak2013,Chau2006,Narayana2011}}, without realizing that such framework is really suitable for canonical SGs. 

Experimentally, a spin glass system is characterized by the temperature dependence of low-field magnetic susceptibility, which shows a sharp peak or cusp at the freezing temperature (T$_f$) {\color{blue}\cite{canmyd1972,Mulder1981}}. In dc magnetization measurements, there are two main protocols to investigate the spin glass systems {\color{blue}\cite{Guy1975}}. First, the zero field cooling (ZFC) protocol where the sample is initially cooled below T$_f$ in absence of magnetic field and the magnetization (M) is measured during warming after applying a small dc magnetic field (H). Another protocol is the field cooling (FC) protocol, where the sample is cooled in the presence of an applied H and the M is measured while warming the sample without changing the field. In the paramagnetic state, magnetization shows identical response namely general Curie-Weiss behavior under both the experimental protocols. However, while the ZFC M vs T curve shows a peak around T$_f$, the FC M vs T curve becomes relatively flat below T$_f$ and bifurcates from the ZFC curve. The magnetization in the ZFC state is strongly time dependent and requires infinite time to reach the equilibrium state below T$_f$. However in the literature the magnetic response in the FC state canonical SG mostly reported to be independent of temperature and time. Therefore, the FC state has been assumed to be the equilibrium state of the system that is equivalent to the ZFC measurements performed over infinite time {\color{blue}\cite{Mydosh2015}}. However, there are some earlier reports showing that this equilibrium description of the FC state may not be totally correct {\color{blue}\cite{Wang, Chamberlin1984, Bouchiat1985,Wenger1984, Nordblad1986}}. Recently, we have reported distinct memory effect in the FC state of canonical spin glass systems reinforcing these earlier claims that the FC state is possibly a non-equilibrium state {\color{blue}\cite{Sudip2020}}. This metastability of the FC state of canonical spin glasses can be considerably different from the metastable behavior observed in the FC state of other kind of glasses, for example, the cluster glass, concentrated spin glass or the glassy state below a kinetically arrested first order transition i.e. magnetic glass. In this context, here we present further evidence of metastability of the FC state of canonical SGs. We also show that the nature of this metastable behavior substantially differs from the well established magnetic response of the FC state of magnetic glasses {\color{blue}\cite{MKC2003, RS2006, KS2006,AB2006, VKS2007, AB2008,PC2008,AB2009,RS2009,RS2013,EPL2013}}. In case of magnetic glass, a system undergoing a first order magnetic transition, may show metastable magnetization at temperatures well below the supercooling limit due to a kinetic arrest of first order phase transition while cooled in certain magnetic field window {\color{blue}\cite{MKC2003, RS2006, KS2006,AB2006, VKS2007, AB2008, RS2013}}. The non-equilibrium nature of a magnetic glass is fairly well known by now, however, such metastable behavior is still described occasionally  in the literature as spin glass like phenomena {\color{blue}\cite{Nayak2013, Chau2006,Narayana2011}}. 

Here we present experimental studies on the non-equilibrium magnetic properties of canonical SG and magnetic glass to highlight: (i) the non-equilibrium nature of the FC state of canonical SG; (ii) the distinct differences between the dynamical magnetic properties of the Canonical SG and magnetic glass systems. In the sections below we present careful magnetic measurements performed on both the ZFC and FC states of two canonical SG systems, AuMn (1.8\%) and AgMn (1.1\%). First we show the presence of  finite thermal hysteresis between the field cooled cooling (FCC) and field cooled warming (FCW) cycles in some temperature range below T$_f$, which to the best of our knowledge has not been reported for any canonical SG system. It underscores the metastable nature of the FC state of canonical SG state. In addition, we have also investigated the effect of thermal cycle on the ZFC and FC state and show that it has quite distinct effects on these states. The metastability of the FC state of canonical SG is further established  with frequency dependence study of ac-susceptibility. While frequency dependence of the ac-susceptibility in the ZFC state is the hallmark of canonical SG, to the best of our knowledge it is the first time the results for such a study on the FC state  of canonical SG is being presented.   Finally, we use a specially designed protocol namely `cooling and heating in unequal field (CHUF)' to probe the non-equilibrium response of canonical SGs AuMn (1.8\%) and AgMn (1.1\%) and magnetic glasses Pr$_{0.5}$Ca$_{0.5}$Mn$_{0.975}$Al$_{0.025}$O$_3$ (PCMAO) and La$_{0.5}$Ca$_{0.5}$MnO$_3$ (LCMO) with contrasting ground states.
 
\section{Experimental details} 
The AuMn (1.8\%) and AgMn (1.1\%) samples are prepared by standard induction melting process. PCMAO has been prepared by standard solid state method and LCMO has been prepared using chemical combustion method. The details of the sample preparation and characterization can be found elsewhere {\color{blue}\cite{PC2008,Nigam1983,SNair}}. The magnetization measurements are performed in MPMS3 SQUID magnetometer, (M/S Quantum design). DC magnetization is measured following three different protocols. In zero field cooled protocol (ZFC), the sample is initially cooled down to T = 2 K from above the freezing temperature (around 5 times T$_f$) in absence of any applied external magnetic field, and then measurements were made while warming the sample in presence of an applied magnetic field. In the field cooled cooling (FCC) protocol, M is measured in presence of a fixed applied field from a temperature greater than T$_f$ to 2 K while cooling down the sample. Then in the subsequent heating cycle,  measurement is made without changing the applied field; this is termed as field cooled warming (FCW) protocol. We also use a specially designed protocol namely `cooling and heating in unequal field (CHUF)', in which magnetization is measured at a fixed measuring magnetic field (H$_M$) during heating after the sample has been cooled every time in a different cooling field (H$_C$) {\color{blue}\cite{AB2006,RS2009}}.  The temperature dependence of magnetization is measured in both temperature stable and sweep mode. For the stable mode, magnetization is recorded after the temperature is made stable. In the sweep mode, the cooling and heating rate is 0.15 K/min. The frequency dependence of ac susceptibility measurements have been carried out in the temperature stable mode.

\begin{figure}[h]
\centering
\includegraphics[scale=0.32]{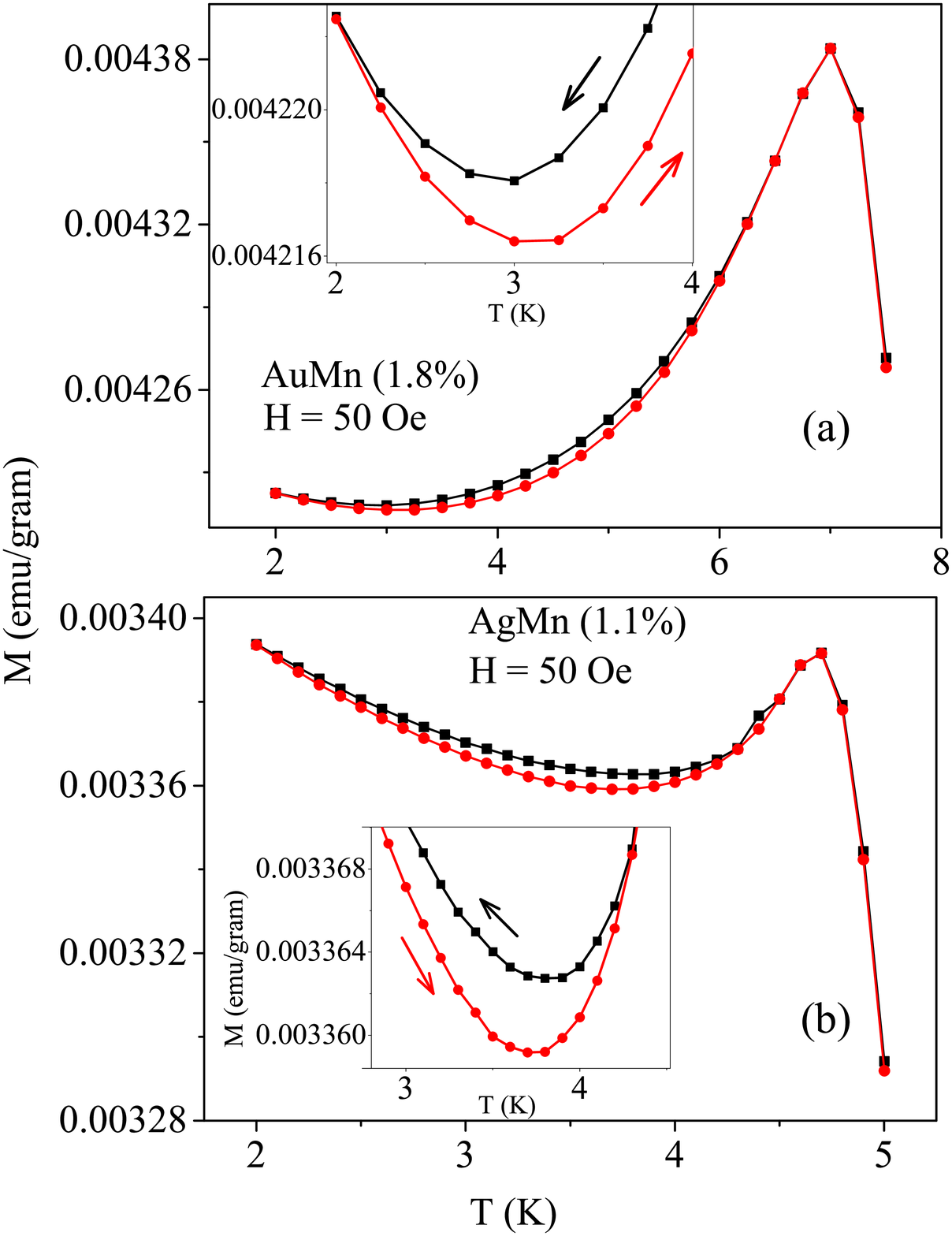}
\caption{ FCC and FCW curve at H= 50 Oe showing thermal hysteresis for AuMn (1.8\%) and AgMn (1.1\%) alloys. The magnetization values at different temperatures has been measured after stabilizing the temperature to reduce the uncertainties in temperature. The arrows in the insets indicate the cooling and heating cycles. }
\end{figure}

\begin{figure}[h]
\centering
\includegraphics[scale=0.32]{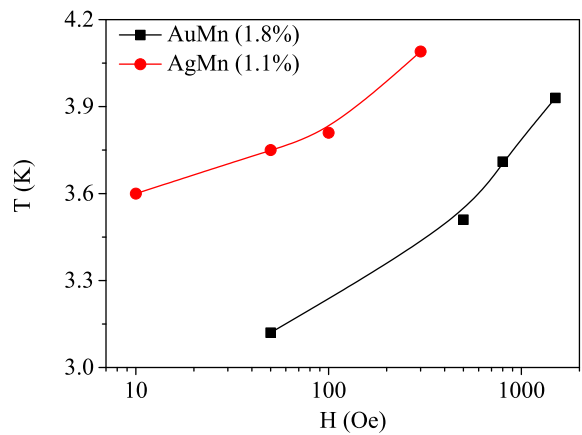}
\caption{ Variation the temperature minima of AuMn (1.8\%) and AgMn (1.1\%) with the applied magnetic field obtained from the FCW cycle (Solid lines are guide to eye).}
\end{figure}

\begin{figure}[h]
\centering
\includegraphics[scale=0.40]{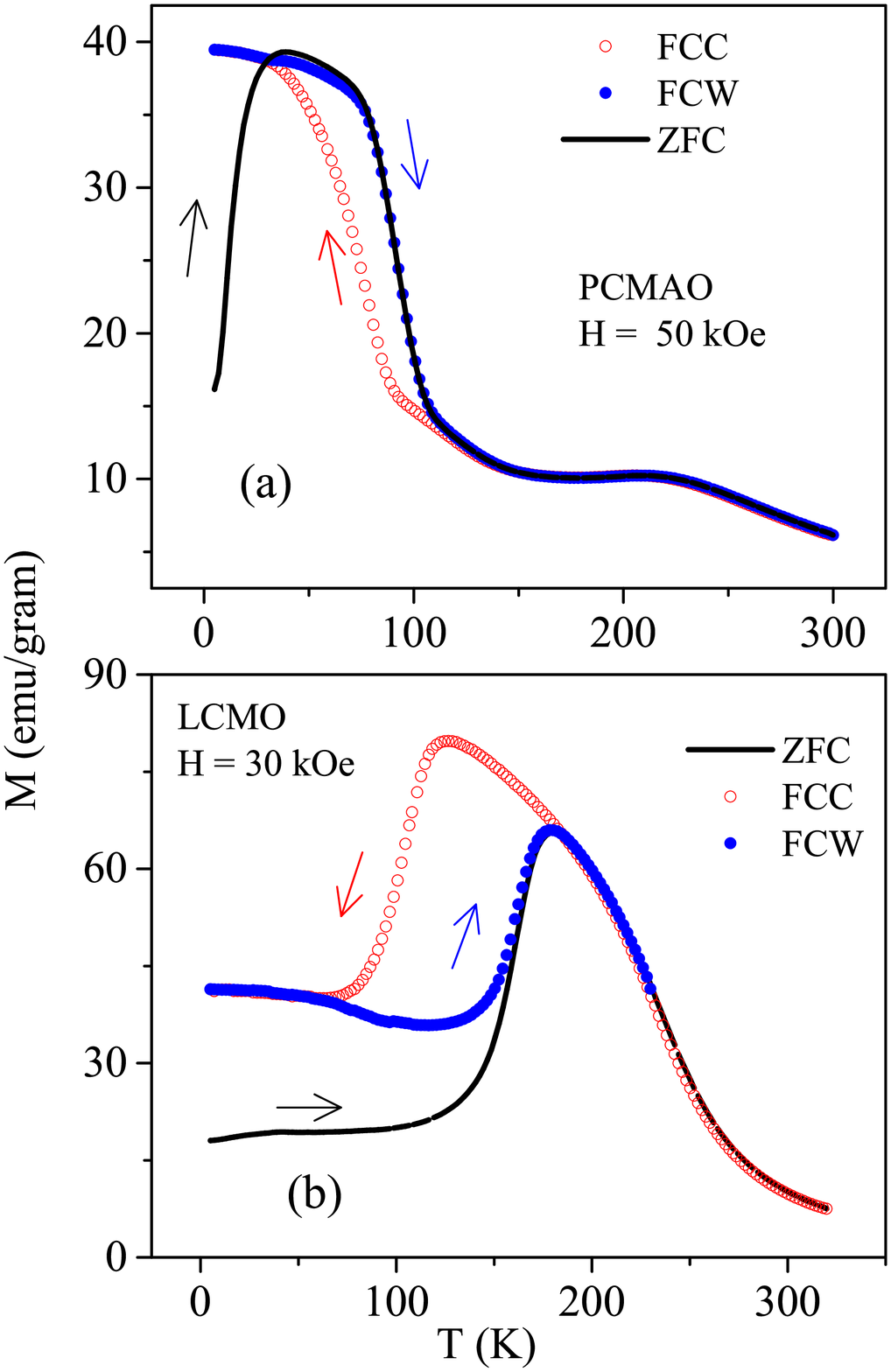}
\caption{ ZFC, FCC and FCW curve of (a) PCMAO measures at H =  50 kOe and (b) LCMO measured at H = 30 kOe. The arrows indicate the direction of the temperature sweep during measurements.}
\end{figure}

\section{Results and discussions} 
\subsection{\label{subsec:level}Thermal hysteresis in the FC state of canonical spin glass and magnetic glass}
The main panel of figure {\color{blue}1(a)} and {\color{blue}(b)} shows the M vs T plots for AuMn (1.8\%) and AgMn (1.1\%) alloys measured in FCC and FCW protocols at the applied field of H = 50 Oe in the stable mode of measurement. Both the curves show a distinct peak at the freezing temperature, T$_f$ = 7 and 4.5 K, respectively, which matches with the earlier reports {\color{blue}\cite{Nigam1983}}. There are few interesting features to be noted carefully in this figure. First, the FC M vs T curves of both the systems show distinct thermal hysteresis  i.e. M$_{FCC} \neq M_{FCW}$ below a temperature, T$_{hys}$, as shown in the insets of figure {\color{blue}1(a)} and {\color{blue}(b)}. This characteristic temperature T$_{hys}$ in both the samples is less than the respective freezing temperature, T$_f$, and the irreversibility temperature, T$_{ir}$, where FC and ZFC M-T curves bifurcate {\color{blue}\cite{Sudip2020}}, i.e. T$_{hys}$ $<$ T$_{ir}$ $<$ T$_f$. Secondly, the FCC and FCW M-T curves show considerable temperature dependence below the freezing temperature, T$_f$. In the FCC protocol, M initially decreases with decreasing temperature, shows a minimum, and then increases again at further lower temperatures. The following FCW curve also shows the minimum. However, there is an interesting difference between AuMn (1.8\%) and AgMn (1.1\%) as shown in the insets of figure {\color{blue}1}. In AuMn (1.8\%) system, the minimum in the FCC M-T curve occurs at T = 0.43 T$_f$, whereas, in AgMn (1.1\%), it occurs around T = 0.78 T$_f$ at the applied field, H = 50 Oe. Moreover, in AuMn (1.8\%) the  FCW M-T curve shows the minimum at higher temperature compared to the FCC M-T curve, and this is opposite to the behavior observed in AgMn (1.1\%). The observed behavior is in contradiction with the common perception that FC magnetization remains flat with the variation of temperature below T$_f$. The FCW magnetization curve always remains below the FCC curves in the hysteresis region in both systems. Thirdly, the minima gradually shifts toward higher temperature as well as it gets suppressed with the increase in applied magnetic field in both the systems. The variation of temperature minima obtained from the FCW curve is shown in figure {\color{blue}2}. It may be noted here that we have performed these measurements in both sweep mode and stable mode and it show similar behavior in both protocols. The observed features in the FC state contradict the common understanding of the equilibrium nature of the FC state in canonical SGs. From the view point of equilibrium state, the FC susceptibility of canonical spin glass is supposed to remain independent of temperature in the mean field scenario {\color{blue}\cite{Parisi2006}} and reversible below T$_f$. The thermal hysteresis between the FCC and FCW curves is not in consonance to the equilibrium picture of the FC state. There are some earlier reports on the dependence of the FC dc susceptibility of spin glass on cooling rate and the presence of thermal hysteresis {\color{blue}\cite{Bouchiat1985,Wenger1984}}, but these aspects have been largely ignored in the literature and the FC state in canonical SG has been continued to be considered as an equilibrium state. We have also measured the FCC and FCW curves at different field. Thermal hysteresis is gradually suppressed with the increase in the value of applied magnetic field. In this context, it may be noted here that the thermal hysteresis is either commonly observed across a first order phase transition due to supercooling and superheating of high temperature and low temperature phase, respectively, or in a non-equilibrium state. It has been proposed earlier that, the FC state in canonical SGs gradually attains the equilibrium state having a lower susceptibility value over the period of long waiting time {\color{blue}\cite{Lundgren1985}}. The observation that FCW magnetization always remains below the FCC magnetization in the temperature region T $< T_{hys}$ (see figure {\color{blue}1}) may be due to the fact that thermal cycle assists the FC state to achieve equilibrium. 

In figure {\color{blue}3(a)} and {\color{blue}(b)}, we have shown the ZFC, FCC and FCW curve of PCMAO and LCMO measured at H = 50 and 30 kOe, respectively. These systems are well studied magnetic glass that is obtained in certain H-T window. PCMAO is paramagnetic at room temperature. As we reduce the temperature, it subsequently undergoes antiferromagnetic (AFM) and ferromagnetic (FM) transition. The AFM to FM transition is a first order phase transition (FOPT), as evident from the thermal hysteresis between FCC and FCW curve. However, when the cooling field  is less than a minimum cutoff field (which is greater than 80 kOe {\color{blue}\cite{AB2009}}) the FOPT is kinetically arrested  and the low temperature state is a mixture of untransformed high temperature AFM phase and the transformed low temperature FM phase. The FM phase is the equilibrium phase and its fraction increases with increasing H {\color{blue}\cite{AB2006, AB2009}}. So, the ZFC state has smaller FM phase fraction compared to the FC state. Therefore, the ZFC magnetization is less than the FCC (or FCW) magnetization as seen in figure {\color{blue}3(a)}. On the other hand, LCMO has opposite phase diagram. It undergoes transition from the paramagnetic state at room temperature to FM state and that is followed by a first order phase transition to AFM state at further lower temperature. Contrary to PCMAO, the ZFC state of LCMO is the equilibrium AFM phase and as the sample is cooled at higher H, the FM to AFM transition gets arrested and the non-equilibrium FM phase fraction increases {\color{blue}\cite{AB2008,PC2008}}.

\begin{figure}[t]
\centering
\includegraphics[scale=0.40]{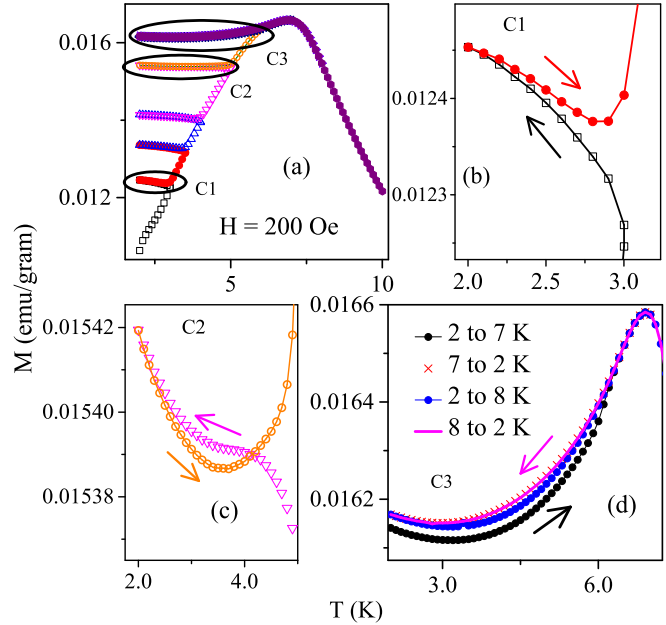}
\caption{ (a) Effect of thermal cycle on the ZFC state of AuMn (1.8\%)  measured in presence of H = 200 Oe performed in sweep mode of the measurement. The sample is first cooled down to T = 2 K at H = 0. Then, H =  200 Oe is applied and the temperature is  raised to progressively higher temperatures and subsequently cooled. The close views of the M-T behavior in the region circled and labelled as C1, C2 and C3 in (a) are shown in (b), (c) and (d), respectively. The thermal cycles are performed in the range (b) 3 to 2 K and back to 3.5 K, in (c) 5 to 2 K to 5.5 K and in (d) 7 to 2 K  to 8 K and subsequent 8 to 2 K and back to 9 K.}
\end{figure}

\begin{figure}[t]
\centering
\includegraphics[scale=0.45]{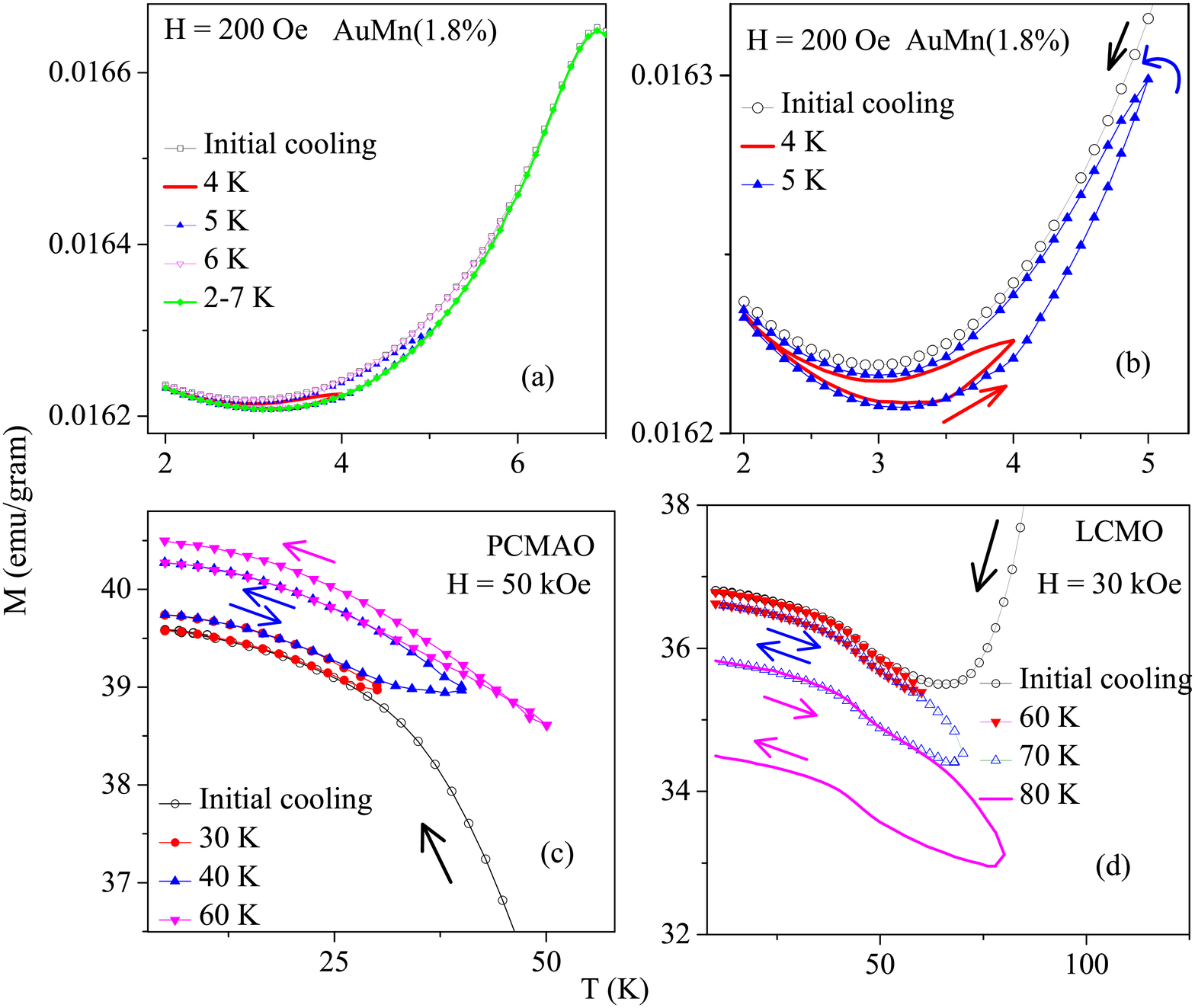}
\caption{ Effect of thermal cycle on the field cooled state of (a) AuMn (1.8\%) at H = 200 Oe, (c) PCMAO at H = 30 kOe and (d) LCMO at H = 50 kOe. Fig (b) presents the magnified view of the temperature dependence of the magnetization during the initial cooling from high temperature and the thermal cycle up to 4 and 5 K of AuMn(1.8\%). In case of spin glass, thermal cycles are similar to minor hysteresis loop. Whereas, in PCMAO, M monotonically increases after each thermal cycle and in LCMO, M monotonically decreases. The arrows indicate the direction of temperature sweep during thermal cycle. The legends indicate the maximum temperature of the thermal cycles after initial cooling in presence of field from room temperature. }
\end{figure}

\subsection{\label{subsec:level}Effect of thermal cycle on the ZFC and FC state of spin glass and magnetic glass:}
To further probe the non-equilibrium nature of the FC state of canonical SG, we have investigated the effect of thermal cycle on the FC state and compared it with the effect of thermal cycle on ZFC state, which is the paradigm of non-equilibrium state. The result for ZFC and FC state of AuMn (1.8\%) have been shown in figure {\color{blue}4} and {\color{blue}5 (a)} respectively. In the ZFC state, we have initially cooled the sample down to T = 2 K and then applied H = 200 Oe. After that we have recorded the M-T curve during warming up to an intermediate temperature, T$_m$ (T$_m$ $<$ T$_f$) and again reduced the temperature to T = 2 K with simultaneously recording the M-T behavior. We have continued similar measurements with progressively higher value of T$_m$ until we reach close to T$_f$. For the FC state, we have performed similar protocol except the initial cooling of the sample, where it is cooled in the presence of field and thermal cycles are performed with different values of T$_m$s. Such thermal cycles is expected to bring a metastable state nearer to the stable state by providing additional thermal energy to the system. We have shown the data in figure {\color{blue}4} and {\color{blue}5(a)} for ZFC and FC state of AuMn(1.8\%) respectively. In case of ZFC state, after each thermal cycle from progressively higher temperature results into a large increase in magnetization and gradually approaching the FC magnetization value. In addition, the thermal cycle affects the magnetization in a very interesting way as we progressively increase T$_m$ toward T$_f$. We have shown  in figure {\color{blue}4(b)}, {\color{blue}4(c)} and {\color{blue}4(d)} the close view of  the M-T data during a couple of thermal cycles those have been marked in the figure {\color{blue}4(a)} as C1, C2 and C3, respectively. The cooling and subsequent heating cycle from all the intermediate temperatures shows irreversibility, pointing to the non-equilibrium nature of the ZFC state. In addition, the nature of this irreversibility also considerably changes at different temperatures, which reveals distinct dynamics in different temperature regime. Finally, when, the thermal cycle is performed within 2 to 8 K and back to 2 K (see figure {\color{blue}4(d)}), it looks similar to the thermal hysteresis observed in the FCC and FCW curve, shown in figure {\color{blue}1(a)}. The thermal cycle, on the other hand, affects the field cooled state quite differently. It looks more like the minor hysteresis loop when returning from progressively higher temperatures, however, the value of magnetization at T = 2 K is found to remain nearly same after each thermal cycle which is equal to the FC magnetization value. This value is, however, is slightly larger than the magnetization value obtained after the thermal cycle from T = 8 K in the ZFC state (see figure {\color{blue}4(d)}).\\ 

In figure {\color{blue}5(c)} and {\color{blue}(d)}, we have presented the effect of thermal cycle in the FC state of  PCMAO and LCMO where the samples have been cooled in presence of H = 50 and 30 kOe respectively from T = 320 K (which is the paramagnetic state in both samples) down to magnetic glass state. It may be noted here that part of this data has been published before {\color{blue}\cite{AB2009}}, but is reproduced here to make the present work self-contained.  In case of PCMAO, at H = 50 kOe the low temperature state is a mixture of equilibrium ferromagnetic  state and non-equilibrium antiferromagnetic state. Thermal cycle from progressively higher temperature at a fixed field deliver thermal energy to the system and the non-equilibrium antiferromagnetic state transforms into ferromagnetic state. This is observed in figure {\color{blue}5(c)} as monotonic increase in magnetization at the low temperature after every thermal cycles {\color{blue}\cite{AB2009}}. On the contrary, in case of LCMO at H = 30 kOe, the thermal cycle at H = 30 kOe transform the ferromagnetic state into antiferromagnetic state and the magnetization decreases after each thermal cycle as shown in figure {\color{blue}5(d)}.

\begin{figure}[t]
\centering
\includegraphics[scale=0.40]{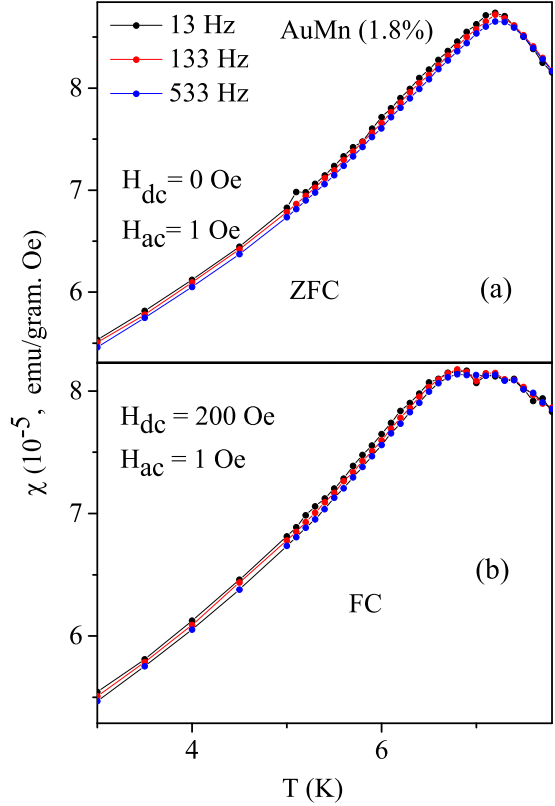}
\caption{ Frequency dependence of ac susceptibility of AuMn (1.8\%)  measured with an ac field of H$_{ac}$ = 1 Oe in: (a) ZFC state and (b) FC state obtained with a cooling field of H$_{dc}$ = 200 Oe .}
\end{figure}

\begin{figure}[h]
\centering
\includegraphics[scale=0.5]{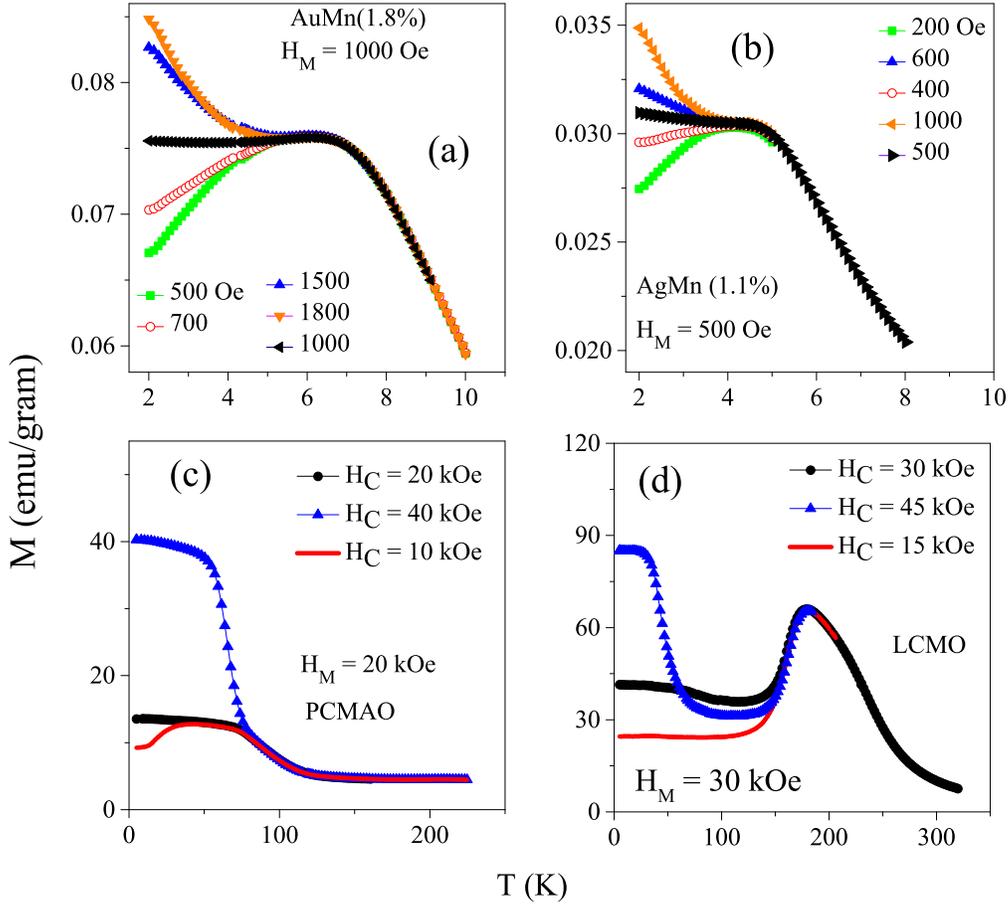}
\caption{ Results of CHUF measurement for (a) AuMn(1.8\%), (b) AgMn(1.1\%), (c) PCMAO and (d) LCMO.}
\end{figure}

\subsection{\label{subsec:level}Frequency dependence of ac susceptibility in the ZFC and FC state of spin glass:}
In figure {\color{blue}6 (a)} and {\color{blue}(b)}, we have shown the ac susceptibility measured at three different frequency in the ZFC and FC state in AuMn (1.8\%) respectively. The frequency dependence of ac susceptibility in the ZFC state has been measured during warming with an ac field of H${ac}$ = 1 Oe after the cooling the sample in zero applied dc field. In the FC state, the sample is cooled down to T = 2 K in presence of dc field, H = 200 Oe and ac susceptibility has been measured during warming in presence of the dc field with H${ac}$ = 1 Oe. The ZFC ac susceptibility shows sharp peak at T$_f$ = 7 K. Within the probing frequency range, the peak temperature increases by a very small amount which is in contrast to the behavior of freezing temperature in concentrated spin glasses, cluster glasses or blocking temperature of superparamagnetic particles where the dependence is larger {\color{blue}\cite{Mydosh2015}}. The frequency dispersion below T$_f$ indicates the broad distribution of relaxation times in the ZFC state of spin glass. It is interesting to note that in FC state, the peak is relatively broad in presence of the dc field, H$_{dc}$ = 200 Oe and it also shows distinct frequency dispersion below T$_f$. However, the peak temperature (T$_f$) has not been observed to get affected significantly. The observed frequency dispersion in the FC state further underlines the non-equilibrium nature of this state, as indicated by the low field dc measurements. 

\subsection{\label{subsec:level}Cooling and Heating in Unequal Field (CHUF) in spin-glass and magnetic glass}
In figure {\color{blue}7}, we have shown the temperature dependence of magnetization measured using a specially designed protocol to probe the non-equilibrium behavior of the field cooled state of a system. In this protocol, magnetization is measured at a fixed measuring magnetic field (H$_M$) during heating after the sample has been cooled every time in a different cooling field (H$_C$) {\color{blue}\cite{AB2006,RS2009}}. This measurement highlights the equilibrium or non-equilibrium nature of the field cooled magnetic state and has been found to be extremely useful earlier in probing the ground state of magnetic glass as will be discussed below. In figure {\color{blue}7(c)}, we have shown the results for PCMAO where H$_M$ = 20 kOe, while H$_C$s are 10, 20 and 40 kOe. As PCMAO is cooled in presence of  H$_C$ = 20 kOe, it results into a kinetic arrest of the  first order transition from antiferromagnetic to ferromagnetic state and produces a low temperature magnetic state which is the mixture of ferromagnetic and antiferromagnetic phase. The most interesting feature of this measurement protocol is that, if a magnetic state is in a state of non-equilibrium, the temperature dependence of magnetization curves at a fixed field, H$_M$ behaves differently depending on whether H$_C$ $>$ H$_M$ or H$_C$ $<$ H$_M$ and it depends on the actual nature of the ground state as well. For example, in case of PCMAO (see figure {\color{blue}7(c)}), although H$_M$ = 20 kOe for every measurement, magnetization at low temperature is different indicating its path dependence. In addition, when H$_C$ = 40 kOe, magnetization is large as compared to the H$_C$ = H$_M$ = 20 kOe curve (reference curve, where H$_C$ = H$_M$) which indicates larger fraction of equilibrium FM phase than reference curve. With increase in temperature, it initially decreases slowly and finally below a certain temperature (around T = 60 K in this case) falls sharply and finally merges with the reference curve. On the other hand, for H$_C$ = 10 kOe, magnetization is smaller than the reference curve which indicates smaller fraction of equilibrium FM state. Magnetization initially increases sharply with temperature and reaches close to the reference curve and becomes flat. Finally again, it decreases sharply and all three curves merges together, highlighting the non-equilibrium state of the low temperature region. Note here that in this case where the ferromagnetic state is the ground state, the M-T curve (for H$_C$ = 10 kOe) changes slope sharply at two temperatures for H$_C$ $<$ H$_M$ and whereas in case of  H$_C$ $>$ H$_M$, the M-T curve shows sharp change in slope only once.

On the other hand, in case of LCMO (see figure {\color{blue}7(d)}), where it undergoes a first order phase transition from high temperature ferromagnetic state to low temperature antiferromagnetic state for low cooling field shows opposite trend. As, we cool the system at higher fields, the transition remains more and more incomplete, increasing non-equilibrium ferromagnetic phase fraction. This is contrary to our previous example where low temperature equilibrium state is ferromagnet and increasing the cooling field increases equilibrium ferromagnetic phase fraction. In case of LCMO, the curve with H$_C$ = 45 kOe ($>$ H$_M$ = 30 kOe) changes slope twice and finally merges with the reference curve (H$_C$ = H$_M$). whereas, for H$_C$ = 15 kOe ($<$ H$_M$) changes slope only once and merges with the reference curve {\color{blue}\cite{AB2008}}. It may be noted here that some part of these data has been published before {\color{blue}\cite{AB2008}}, but is reproduced here to make the present work self-contained.

We have applied similar protocol for canonical spin glasses. The data have been shown in figure {\color{blue}7(a)} and {\color{blue}7(b)} for AuMn(1.8\%) and AgMn(1.1\%) samples respectively. In the case of AuMn(1.8\%), the measuring field H$_M$ = 1000 Oe and different cooling fields produce different magnetization curves. For the H$_C$ $<$ H$_M$, the magnetization at 2 K starts at lower value as compared to the magnetization for  H$_C$ = H$_M$. For  H$_C$ $>$ H$_M$, the trend is opposite. As, the temperature is increased, all the curves merge with the reference curve, H$_C$ = H$_M$ = 1000 Oe at a temperature, T$_{merge}$ which is smaller than T$_f$. In contrast to the cases of magnetic glasses, the M - T curves do not show any sharp change in slope in canonical SGs.
 
\section{Summary and conclusion} 
Summarizing we can say that we have investigated the magnetic response of the FC state of two representative canonical SG systems AuMn(1.8\%) and AgMn(1.1\%)  in details. In combination with our recent study of the memory effects in the same canonical SG systems $\color{blue}\cite{Sudip2020}$ the results of the present study unequivocally establish the distinct non-equilibrium nature of the FC state of the canonical SGs. The characteristic features of this non-equilibrium response of the FC state is quite different from those of the ZFC state, and this difference has been highlighted. There has been some earlier suggestions $\color{blue}\cite{Wang,Chamberlin1984,Bouchiat1985, Wenger1984,Nordblad1986}$ that the FC state of canonical SGs may not be an equilibrium state. However, over the years FC state of the canonical SGs has not been subject of intense scrutiny as much as the ZFC state, presumably due to the popularity and the implicit acceptance of the thermodynamic phase transition picture of the canonical SG state. Our present studies along with these earlier results clearly indicate the presence of a rugged energy landscape in the FC state of canonical SGs. In this direction there exists some suggestion of exponentially increasing `sparsity' of thermally accessible independent free energy levels with the decrease of temperature in the SG state $\color{blue}\cite{hooger1985}$.

Furthermore we have compared the non-equilibrium response of the FC state of canonical SG with that of the two representative magnetic glass systems namely PCMAO and LCMO. The distinct differences in the non-equilibrium properties of these two classes of magnetic materials is clearly distinguished. In the literature there is a tendency to attribute any non-equilibrium response observed in magnetic materials to spin-glass behavior. The present study indicates that each class of magnetic systems has its own finger-print magnetic response, which can be easily identified with simple but careful experiments. 

\section{\label{sec:level}References: }  

\end{document}